\documentclass{aastex63}
\usepackage{amsmath}
\received{xxx}
\revised{xxx}
\accepted{xxx}
\submitjournal{ApJ}

\shorttitle{Magnetic effect on the rotational frequency of a neutron star}
\shortauthors{F\"ahnle and Wunner}
\begin{document}

\title{Magnetic Effect on the Rotation Frequency of a Neutron Star}

\correspondingauthor{M. F\"ahnle}
\email{faehnlemanfred2704@gmail.com, guenter.wunner@itp1.unistuttgart.de}

%\collaboration{1}{(AAS Journals Data Scientists collaboration)}

\author{M. F\"ahnle}
\affiliation{Formerly with the Max-Planck Institute for Intelligent Systems Stuttgart,
\\
Sch\"onblickstra\ss e 95
71272 Renningen, Germany}

\author{G. Wunner}
\affiliation{University of Stuttgart, Institute for Theoretical Physics 1, \\
  Pfaffenwaldring 57, 70569 Stuttgart, Germany}



\begin{abstract}

  A neutron star contains regions with particles which exhibit magnetic moments, which, in turn,  generate respective magnetizations. By magnetoelastic interactions between the magnetization and the matter there arises a magnetostriction. The magnetostriction changes the mass density of the neutron star and hence its inertia tensor, and as a result it changes the rotation frequency of the star.
We also sketch how the calculation of the magnetostriction
in the different inner regions of the neutron star has to proceed, and which parameter inputs are needed. 
\end{abstract}



\section{Introduction} \label{sec:intro}

In this paper we point out the effect of magnetostrictions in the interior of a neutron star on its rotation frequency. The exact composition of the interior of a neutron star still is an ongoing debate, since it critically depends on the equations of state for the different constituents of the interior used in calculations (for reviews see, e.\,g., \cite{baym18}, \cite{lattimer07}, or \cite{jiang19}, and references therein). Further progress is expected from the Neutron Star Interior Composition Explorer (NICER) (\cite{bogdanov19a, bogdanov19b}) installed on the International Space Station which is devoted to the study of neutron stars through soft X-ray timing. 

To illustrate the magnetostrictive effect on the rotation frequency of a neutron star, and be specific, we resort to a simplified model for the interior, presented by \cite{ruderman98}. In this model the 
neutron star is composed of different regions with different properties. In the interior there is assumed a liquid state,
with neutrons which exhibit superfluid properties, and with protons which form a type-II superconductor with superconducting flux lines carrying magnetic moments (\cite{srinivasan90}). The crust of the star is assumed as a  solid, and the outermost shell of the crust  is formed 
by completely ionized iron atoms which carry nuclear magnetic moments, and a sea of relativistc electrons which carry electronic magnetic moments (\cite{chamel08}).

Usually, a magnetic effect on the mass density of the neutron star, which determines the inertia tensor, is not taken into account in the literature. Here we show that such an effect exists, and that this affects the rotational frequency of the star.

It is worthwhile mentioning that in general the rotation frequency of the neutron star itself is not a constant. There are physical processes which accelerate the rotation (spin-up) and physical processes which retard the rotation (spin-down). An example for a spin-down process is the emission of electromagnetic waves by which the star loses energy.  The radiation originates in the magnetosphere of the neutron star, and is magnetic dipole radiation, because in general the magnetic field axis of the neutron star is not aligned with its rotation axis.

\section{Magnetostriction in the Neutron Star} \label{sec:magnetostriction}

The magnetic moments of the particles in the various regions of the star discussed in Section \ref{sec:intro} produce  magnetizations in those regions. Due to the magnetoelastic interactions of the magnetization with the matter,  magnetostrictive strains arise which change the mass density of the star and, thus, the inertia tensor of the neutron star. From this it follows that there is a magnetic effect on the rotation frequency of the star.

In this Section we want to sketch how the calculation of the magnetostriction
in the different inner regions of the neutron star would have to proceed, and which parameter inputs are needed. The underlying theory is the micromagnetic theory described in \cite{kronmueller03}. The basic quantitity entering the calculation of the magnetostriction is the so-called quasiplastic strain tensor $\boldsymbol{\epsilon}^{Q}$.

An expression for the components of this tensor in the isotropic liquid interior of the star is given in  \cite{hubert98},
\begin{equation}
 \epsilon^{Q}_{ik} = \frac{1}{2} \lambda_s (\gamma_i \gamma_k - \frac{1}{2} \delta_{ik}).
  \label{eq1}
\end{equation}
Here $\lambda_s$ is a constant which characterizes the magnetostrictive properties of the isotropic liquid. The $\gamma_{i}$ are the direction cosines of the magnetization, which is assumed homogeneous in the liquid part, not position-dependent. 
The quantity $\delta_{ik}$ represents the usual Kronecker symbol.

An expression for the components of the tensor in the anisotropic solid crust of the star is also given in \cite{hubert98},
\begin{eqnarray}
\epsilon^{Q}_{ik} &=& \frac{3}{2} \lambda_{100} (\gamma_i^2-\frac{1}{3}) \quad \hbox{for} \quad i=k=1,2,3 \nonumber \\
  \epsilon^{Q}_{ik} &=& \frac{3}{2} \lambda_{111} \gamma_i \gamma_k \quad \hbox{for} \quad  i  \neq k.
                        \label{eq 2}
\end{eqnarray}
The quantities $\lambda_{100}$ and $\lambda_{111}$ are magnetostriction constants which correspond to the fractional changes of the length of the sample upon saturation in [100]- direction and in [111]-direction, respectively.

The magnetostrictive strains of the two regions add up to a total strain.
In the following the magnetostrictions in the liquid and in the solid region of the neutron star are calculated separately.

\subsection{Isotropic liquid interior}

In the liquid part the elastic (magnetostrictive) strain $\boldsymbol{\epsilon}^{el}$
is determined by minimizing 
the elastic potential $\Phi^{el}$ given in \cite{kronmueller03} with respect to $\boldsymbol{\epsilon}^{el}$,
\begin{equation}
\Phi^{el} = -\frac{1}{2} \int (\boldsymbol{\epsilon}^{Q}\cdot \cdot ~\boldsymbol{C} \cdot \cdot ~\boldsymbol{\epsilon}^{Q}-
\boldsymbol{\epsilon}^{el} \cdot \cdot ~\boldsymbol{C} \cdot \cdot ~\boldsymbol{\epsilon}^{el}) d^3 \boldsymbol{r}\,,
\label{eq3}
\end{equation}
In (3) the symbol $\cdot \cdot$ denotes the tensor product.
This potential includes the quasiplastic strain tensor $\boldsymbol{\epsilon}^{Q}$, 
the tensor $\boldsymbol{C}$ of elastic constants, as well as the tensor $\boldsymbol{\epsilon}^{el}$.   The magnetization does not depend on the position, and therefore the integrand does not depend on position either. Thus the integral is just the integrand times the volume of the liquid region.

To calculate the value of the integral one has to insert the respective components of the tensor of elastic constants for the isotropic liquid and the volume of this region. Furthermore, one has to insert a value for the magnetization in the interior part of the neutron star arising from the superconducting flux tubes. For different neutron stars the magnetizations and the volumes of the various regions may be very different. This is a consequence of the fact that the magnetic fields of neutron stars lie in the range from $10^8$ Gauss up to $10^{15}$ Gauss \cite{reisenegger01}.

\subsection{Crust}

In the solid outermost shell of the crust the calculation of the magnetostriction follows a different but similar line.  As shown in \cite{pethick98}, the elastic deformation energy, which represents $\Phi^{el}$,  cannot be represented by strains, but the fundamental variable rather is the displacement field $\boldsymbol{u}$  of the atoms. From $\boldsymbol{u}$ the strain tensor  $\boldsymbol{\epsilon}$ can be calculated via
\begin{equation}
\boldsymbol{\epsilon} = \frac{1}{2} \left(\nabla \boldsymbol{u} + \nabla \boldsymbol{u}^t \right),
\end{equation}
where the superscript $t$ denotes transposition.

Inversely, the displacement field $\boldsymbol{u}$ can be calculated also with this equation for given strain tensors $\boldsymbol{\epsilon}$. From a given quasiplastic strain tensor $\boldsymbol{\epsilon}^Q$ the corresponding quasiplastic displacement field $\boldsymbol{u}^Q$  can be calculated, and for given elastic strain tensor $\boldsymbol{\epsilon}^{el}$ the corresponding elastic displacement field $\boldsymbol{u}^{el}$ can be calculated.

The elastic potential $\Phi^{el}$ given by eq. (3) for the liquid region, describes the elastic deformation energy of the system. The first part is the energy resulting from the quasiplastic strains $\boldsymbol{\epsilon}^Q$, the second part is the energy resulting from the elastic strains $\boldsymbol{\epsilon}^{el}$. The elastic potential for the solid shell is the counterpart to the elastic potential of the liquid region. The two parts of the elastic potential $\Phi^{el}$ for the solid part again describe the elastic deformation energy, the first part the energy related to the quasiplastic displacement field $\boldsymbol{u}^Q$, the second part the energy related to the elastic displacement field $\boldsymbol{u}^{el}$. The quasiplastic displacement field $\boldsymbol{u}^Q$   is calculated via eq. (4) from the quasiplastic strain field $\boldsymbol{\epsilon}^Q$, and the  elastic displacement field $\boldsymbol{u}^{el}$ is determined by minimizing the following elastic potential $\Phi^{el}$ of the nonisotropic shell of the crust  with respect to $\boldsymbol{u}^{el}$.

For the elastic deformation energy in the outermost shell of the neutron star 
we can use eq. 11 of \cite{pethick98}. In this way we obtain
\begin{eqnarray}
  \Phi^{el} &=& \ - \frac{1}{2} \int  \Bigg\{ - \frac{B}{2} \left( \frac{\partial u_{el,x}}{\partial x} + \frac{\partial u_{el,y}}{\partial y} \right)^2 + \frac{C}{2}
                \left[\left(\frac{\partial u_{el,x}}{\partial x} - \frac{\partial u_{el,y}}{\partial y} \right)^2 + \left( \frac{\partial u_{el,x}}{\partial x} + \frac{\partial u_{el,y}}{\partial y} \right)^2    \right] \nonumber \\
            &+& \frac{K_3}{2} \left(\frac{\partial^2 u}{\partial z^2} \right)^2 + B^\prime \left(\frac{\partial u_{el,x}}{\partial x} + \frac{\partial u_{el,y}}{\partial y}     \right) \left( \frac{\partial u}{\partial z} \right)^2
+\frac{B^{\prime\prime}}{2}  \left( \frac{\partial u}{\partial z} \right)^4  
                                \nonumber \\[1.7ex]
  &~& \hskip 2 truecm \hbox{plus the same sum of terms with~} \boldsymbol{u}^Q \hbox{~instead of~}\boldsymbol{u}^{el} \Bigg\} \; d^3\boldsymbol{r} \, .
\end{eqnarray}  

%\section{Change of the Inertia Tensor} \label{sec:inertia_tensor}

\section{
{\bf   Input necessary to calculate the size of the magnetic effect}}
\label{sec:estimate}
The magnetostriction changes the mass density of the star. The inertia tensor of the spherical star may be calculated from the mass densities in the various regions. To do this, one has to start from the initial mass density, which then is changed by the magnetostrictive strains. Because the rotation frequency of the star depends on the inertia tensor and on the angular momentum of the matter of the neutron star, the change of the initial mass density due to the magnetostrictive strains results in a magnetic effect on the rotational frequency.

{\bf We now discuss the various inputs which are needed to calculate the size  of the magnetic effect. To start with one} must insert values for the magnetizations in the various regions of the neutron star, because these magnetizations determine the quasiplastic strains $\boldsymbol{\epsilon}^Q$  and the quasiplastic displacement field $\boldsymbol{u}^Q$ which enter equations (3) and (5).  In the envelope of the solid crust there are completely ionized Fe atoms which carry a  nuclear magnetic moment, and   a sea of relativistic electrons which carry an electronic magnetic moment, $m_{e}$, respectively. Because the nuclear moment is much smaller than the electronic moment, {\bf one} has to take into account only the electronic moments. The magnetization then is $M_{el} =  Z m_{el} \varrho$, where $\varrho$ is the density of Fe atoms  and $Z$ is the total number of electrons in the Fe atom, given by the nuclear charge number $Z =26$. Inserting an electron number density $\varrho$ of $10^4$ per cm$^3$, this yields a magnetization of about $242 \times 10^{-14}$ A/m.

The magnetization of the liquid superconducting proton region with flux tubes  is $M=m_{\rm tube}  den_{\rm tube}$, where $m_{\rm tube}$ is the magnetic moment of a single tube and $den_{\rm tube}$  is the density (number of tubes per area) of the flux tubes. The change of the total magnetic moment $m$ of the sample by the appearance of $n$ tubes is then $\Delta m = n \Phi_0/\mu_0$, with the elementary flux quantum $\Phi_0 = 2.07 \times 10^{-7}$ Gauss/m$^2$  and with $\mu_0 =  4 \pi \; 10^{-7}$ Vs/Am. From this the dipole moment of the flux tube can be calculated. The area density of the flux tubes  is (cf. \cite{ruderman98})  $den_{tube} \approx 10^4 /(P(s)\, \hbox{cm}^2)$, where $P(s)$ is the period of the vortex rotation in units of seconds.

{\bf An anonymous referee points out to us that our estimates of the magnetizations in the various parts of the neutron star are based on the assumption that there is a complete alignment of the respective magnetic moments. In fact, the magnetic moments related to
the flux lines of the type-II superconductor in the liquid interior of the neutron star  may deviate from a complete alignment because of the rotation of the star. The magnetic moments of the relativistic electrons in the solid outer crust of the neutron star also may not be completely aligned because of thermal disorder. Furthermore, the referee notes that in neutron stars there may be a contribution to the magnetization generated by macroscopic currents, which is correct. In our theory only magnetizations appear, independent of the physical processes which generate them.}

The constants which appear in the elastic deformation energy of the non-isotropic solid envelope of the crust (which appear in eq. (5)) are given in \cite{pethick98}, the values of the tensor of elastic constants for the liquid isotropic region of the interior of the star are yet unknown. Even if {\bf one} inserted reasonable values for them, a very big problem for the calculation of the magnetic effect remains, because the magnetostriction constant $\Lambda_s$ of the isotropic liquid region is not known, and also the magnetostriction constants $\Lambda_{100}$  and $\Lambda_{111}$ of the non-isotropic solid envelope of the crust are completely unknown.

It is true that for some transition metals and for some intermetallic compounds
the values of these constants are known (see Table 2.3 of \cite{kronmueller03})  at temperatures of 4.2 K and at room temperature, and for densities of these materials which they have on the surface of the earth. But in the liquid region and in the solid envelope of the crust of the neutron star the temperatures and the densities are drastically higher, so that it does not make sense to insert the values given in that table of \cite{kronmueller03}.

Therefore, a {\bf final calculation} of the magnitude of the magnetic effect on the rotation frequency of the neutron star must be postponend to the future, when more information, either theoretical or experimental, on these constants has become available. But the main message of this paper remains that it is important to point out the existence of such a magnetic effect.
 
\section{Summary}\label{sec:summary}
In the present paper we have pointed out a magnetic effect on the rotation frequency of a neutron star. To the best of our knowledge this magnetic effect has  not been taken into account in the literature so far. It results from the magnetoelastic interaction of the magnetizations in the various regions of the neutron star with matter. This interaction generates magnetostrictive strains, which change the mass density of the star and, thus, the inertia tensor. Because the rotation frequency of the star depends on the inertia tensor and on the angular momentum of the star material, this produces a change in the rotational frequency of the neutron star. All this is an interesting combination of the knowledge of the astrophysical properties of the neutron star with the knowledge of how magnetostricton is produced in a magnetic material.

{\bf \cite{easson77} proposed an anisotropic stress tensor  for the type-II proton superconductor in neutron stars, and the question arises whether this is equivalent to the magnetostriction effects discussed in the present paper. Note, however, that  the magnetostriction generates  an anisotropic strain of the material by the magnetoelastic interaction of the magnetization with the matter, i.e., in the present paper the magnetostrictive strains are investigated, rather than the stress produced by the magnetostriction. All quantities appearing in the magnetostriction formalism are strains and not stresses. Therefore the discussion by \cite{easson77} is not equivalent to the discussion in the present paper. Furthermore, the anisotropic stress  is predicted only for the liquid interior of the neutron star, whereas in the present paper both the liquid interior and the solid crust envelope are taken into account.

The question also arises whether the magnetic effect on the rotation frequency could be detected by observing the change of the rotation frequency in time. Such a change due to a time-dependence of the magnetic effect would appear when the magnetizations and hence the magnetostrictions in the various parts of the neutron star themselves would vary. In the liquid inner part of the neutron star this could happen as a result of the rotation of the star. In the solid outer crust of the star the change could appear because the charged particles are accelerated or retarded due to the spin-up and spin-down processes. However, the change of the magnetic effect on the rotation frequency  caused by these changes of the magnetizations and hence of the magnetostrictions in the various parts of the star could hardly be distinguished from the changes of the rotation frequenciy by the above mentioned spin-up and spin-down processes.} 

\acknowledgements

We thank Malvin Ruderman and Chris Pethick for helpful correspondence.

%\bibliography{refs}{}
%\bibliographystyle{aasjournal}



\end{document}